\documentstyle[11pt,moriond,epsfig]{article}

\def\be{\begin{equation}}
\def\ee{\end{equation}}
\def\bea{\begin{eqnarray}}
\def\eea{\end{eqnarray}}

\begin{document}
\vspace*{4cm}
\title{AMS : A PARTICLE OBSERVATORY IN SPACE}

\author{ Aur\'elien Barrau}

\address{Institut des Sciences Nuc\'eaires\\
53, av des Martyrs, 38026 Grenoble cedex, France\\
Universit\'e Joseph Fourier, 38400 St Martin d'H\`eres, France\\
{\rm AMS collaboration}}

\maketitle

\abstracts{The expected physics and astrophysics capabilities of the AMS experiment on 
board the International Space Station are briefly reviewed.}

\section{Introduction}

The AMS spectrometer will be implemented on the International Space Station at the
end of 2003 (or beginning of 2004). The instrument will be made of a superconducting 
magnet which inner volume 
will be mapped with a tracker consisting of 8 planes of silicon microstrips surrounded 
by several sub-detectors: a time-of-flight (TOF), a ring imaging Cherenkov (RICH),
an electromagnetic calorimeter (ECAL), a transition radiation detector (TRD) and, possibly, a
synchrotron radiation detector (SRD). The expected performances will be reviewed
together with the physics goals : high statistics study of cosmic rays in a wide
range of energies, search for primordial antimatter, search for non-baryonic dark
matter and, to some extent, gamma-ray astrophysics.

A simplified version of the spectrometer has already been flown during a precursor
flight (june 2-12, 1998) on the Space Shuttle Discovery. The technical success
together with the first physics results and their interpretation will be briefly
reviewed.

\section{The spectrometer}

\begin{figure}
\begin{center}
\psfig{figure=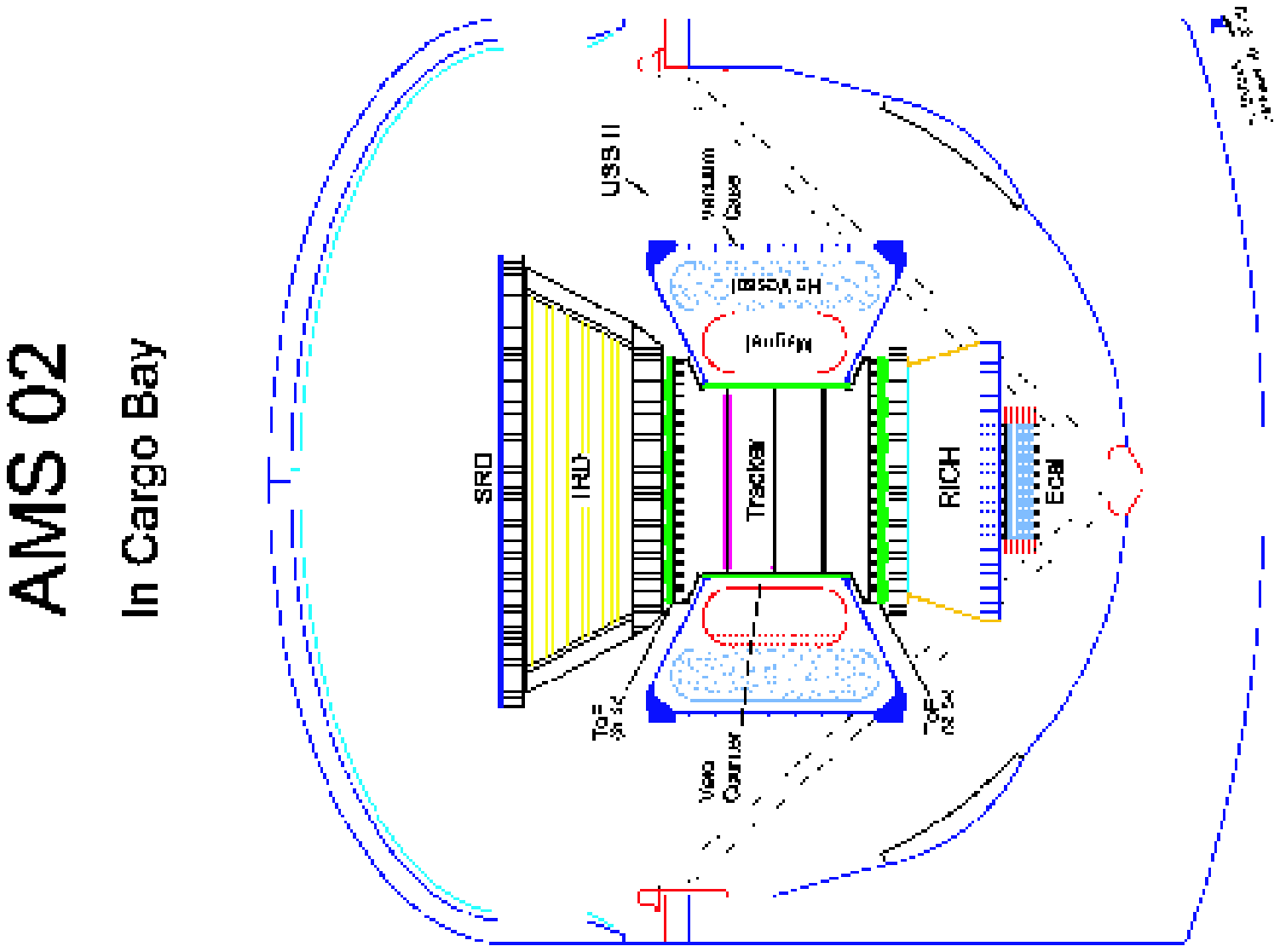,height=7cm,angle=-90}
\caption{the AMS spectrometer (to be implemented on the ISS)
\label{fig:AMS}}
\end{center}
\end{figure}

The AMS spectrometer Fig.\ref{fig:AMS} will be based on a superconducting magnet which maximum field should
be of the order of 1 Tesla, perpendicular to the axis of the cylinder. Its inner
volume will be filled with 8 layers of double sided silicon tracker measuring the 
trajectory of the charged particles and, therefore, their rigidity (from around 300~MV up to
3~TV). The time of flight system has its planes at each end of the magnet, covering the outer
tracker layers. It provides information on the particles transit time and will be used
both as a trigger for the whole experiment and as a measurement of the velocity for low
energy events. The transition radiation detector will be useful to improve the lepton/hadron
discrimination up to the proton threshold around 300~GeV.
A three-dimensional sampling electromagnetic calorimeter will measure $\gamma$, $e^+$ and
$e^-$ energies and increase the lepton/hadron rejection factor to $10^3$ (on a limited
fraction on the acceptance). Finally, the ring imaging Cherenkov detector will allow precise measurements of the velocity $\Delta
\beta / \beta \approx 10^{-3}$ together with a charge determination up to $Z \approx 20$.

\section{Cosmic-ray astrophysics}

Cosmic rays are a part of the galactic components which reflects the dynamical equilibrium of the
physical system they belong to, carrying out very important informations on the galactic medium,
its population, structure, and history. In particular, it has been recently shown \cite{maurin}
that measurements of cosmic nuclei give strong constraints on the diffusion parameters and
rule out some models ({\it e.g.} those without convection or reacceleration and Kolmogorov
spectra for the diffusion coefficient).

The AMS experiment will dramatically improve the situation, allowing a high statistics study of many cosmic ray species including $e^+$,
 $e^-$, $p$, $\bar{p}$, and the lightest ions $d,t,^{3,4}$He. Heavier light ions will also 
be studied with mass identification up to A$\approx 20$, and elements up to Z$\approx 20$ 
depending on the final performances of the instrumentation of the spectrometer (RICH in 
particular).
Unstable ions like $^{10}$Be, and $^{26}$Al are of particular interest since they provide
a measurement of the time of confinement of charged particles in the galaxy (galactic 
chronometers) \cite{BO00}. 
The corresponding antimatter nuclei will be searched with equivalent instrumental 
performances in identification and kinematic range. 
The metrological perspectives are summarized in table~\ref{PROSP} borrowed from
Buenerd, 2001 \cite{Bu}.

\begin{figure}
\begin{center}
\psfig{figure=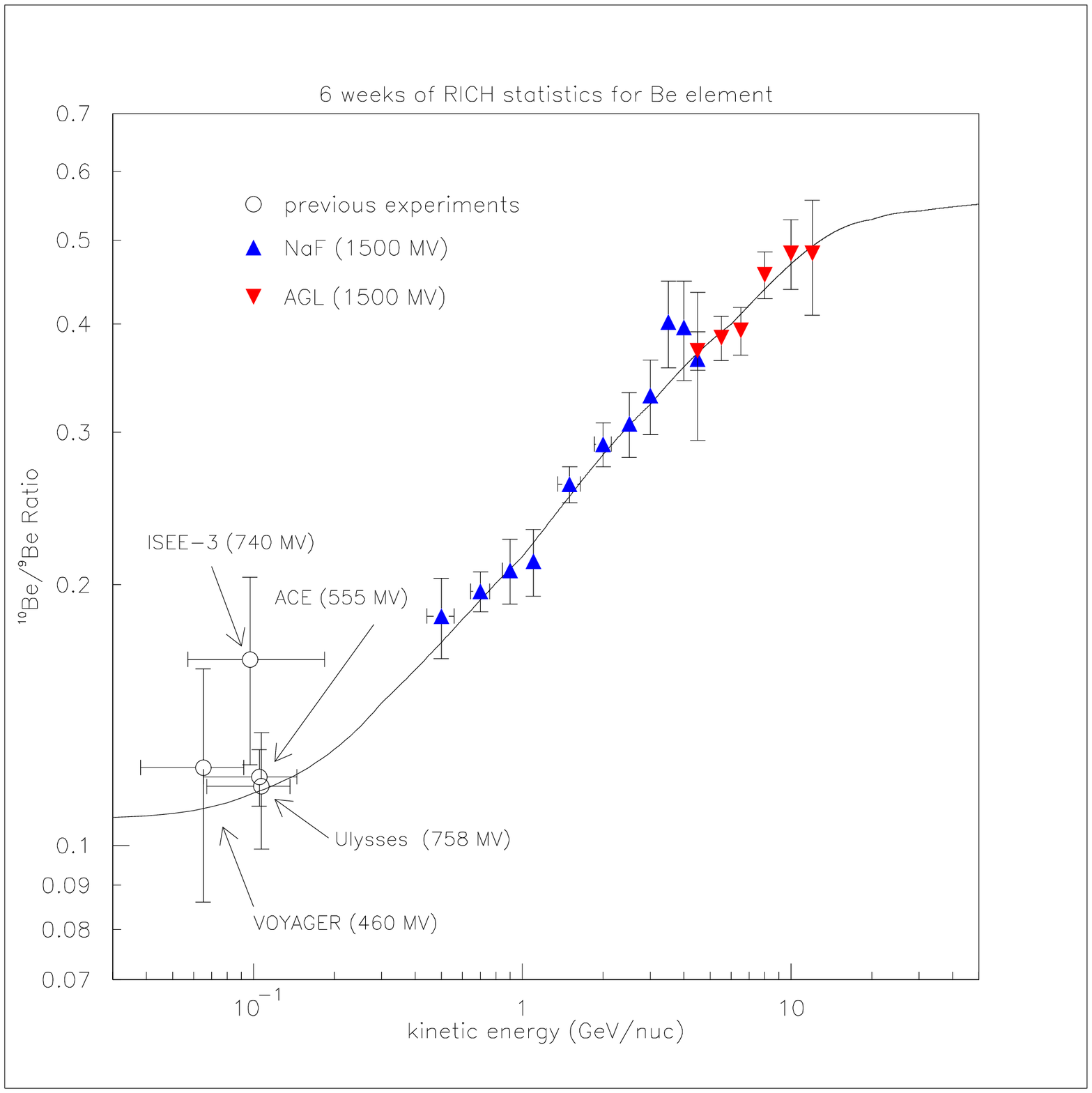,height=7cm}
\caption{
\label{fig:alexi}Expected statistics for the $^{10}$Be isotope for 6 weeks of counting 
with AMS.}
\end{center}
\end{figure}

Basically
the spectrometer will be able to accumulate statistics larger by 3 to 4 orders of magnitude
than those measured so far by other space-borne experiments, for all the species studied. Figure
\ref{fig:alexi} gives the expected statistics for the $^{10}$Be isotope for 6 weeks of counting 
from a simulation using the RICH of AMS and the previous measurements (points of the
lower left edge of the plot) \cite{BO00}.

\begin{table}
\label{CHARAC}
\begin{center}
\vspace{0.5cm}
\begin{tabular}{llll}
\hline\hline
 Particles        & P$_{min}$ & P$_{max}$ & Comments  \\
\hline 
e$^-$             & $\approx$0.3 & $\approx$3000 & Upper limit set by rigidity resolution \\
e$^+$             & $\approx$0.3 & $\approx$300  & Upper limit set by TRD \\
proton            & $\approx$0.3 & $\approx$3000 & Upper limit set by rigidity resolution \\
\hline
\multicolumn{4}{c}{\bf Charge identification of elements } \\
Ions Z$<\approx$20 & $\approx$0.3 & $\approx$1500  & set by RICH performances \\
\hline
\multicolumn{4}{c}{\bf Mass identification of isotopes} \\
Ions A$<$4             & 1 to 4  & $\approx$20 & set by RICH performances \\
Ions 4$<A<\approx$20 & 1 to 4  & $\approx$12 & \\
\hline
\multicolumn{4}{c}{\bf Antimatter } \\
$\bar{p}$         & $\approx$0.3 & $\approx$3000 & Depending on $\bar{p}/e^-$ discrimination\\
$\overline{ions}$ & $\approx$0.3 & $\approx$1500 & $\overline{He}$, $\overline{C}$ \\
\hline \hline 
\end{tabular}
\caption{\it Summary of the particle detection and identification range of the AMS02 
 spectrometer. The upper instrumental limits are set either by the momentum measurement 
 accuracy (at the highest momenta) or by the range of identification of the particle. The 
 lower values are set by the low momentum cutoff of the magnetic spectrometer or by the 
 range of particles in detectors, or by threshold effects. Statistical 
 limits are ignored. The given numbers should be considered as orders of magnitudes.
 The true limits will depend critically on the relative statistics of the particles to be 
 identified versus background particles, like e$^+$/$p$ or $\bar{p}$/$e^-$.  
 The momenta are given in GeV/c or GeV/c per nucleon when applicable.} 
\label{PROSP}
\end{center}

\end{table}

\section{Search for dark-matter and "new-physics"}

It is now well known that stars are less than 1\% of the mass-energy content of the Universe.
Baryons, whose density can be inferred from deuterium abundance measurements in high-redshift
hydrogen clouds within the Big-Bang nucleosynthesis model, should contribute around 5\% (this
value is in agreement with several other independent estimates). Non-baryonic dark matter
can be measured in rich clusters, comparing the total mass (given by the motion of galaxies or
by weak lensing) with the gas mass (given by X-ray flux or by Sunyaev-Zel'dovich CMB
distortion). The resulting density is around 40\% (which is consistent with a variety of other
methods involving different physics). As a conclusion, the non-baryonic content of the Universe
seems to be much more important than the baryonic one \cite{turner}. One of the best candidates 
for this unknown component would be long-lived, weak-interacting massive particles : exactly
what could be expected from relic neutralinos if R-parity is conserved. AMS will look for dark
matter through the annihilation of neutralinos, expected to "enrich" the $\bar{p}$ and e$^+$ 
spectra. Recent studies have shown that the $\bar{p}$ flux should, unfortunately, not be
substancially distorted by this source-term, due to non-annihilating inelastic scattering of
secondary antiprotons and to nuclei-induced antiproton creation \cite{simon} \cite{Ber}. On the
other hand, the secondary-antiproton spectrum is now computed with great accuracy \cite{donato},
allowing a good sensitivity to the absolute normalisation of the spectrum and not
only to its shape. Stringent upper limits on some supersymmetric models (if not a positive
detection) could, this way, be derived with AMS-data. Furthermore, antideuterons \cite{donato2}
have also been shown to be a powerful tool to study neutralinos annihilation (due to a tiny
secondary cosmic background, mostly for kinematical reasons) if the instrumental background (due do
mis-reconstructed deuterons and anti-protons events) can be lowered enough. This latter point is
quite unlikely. Finally some "exotic" astrophysical objects, like primordial black holes 
\cite{jane}, can be very efficiently looked for through the antiprotons emission 
\cite{orito}.

From a technical point of view, it will also be extremely useful to have the same detector
measuring, for the first time, both the cosmic nuclei (to constrain the diffusion model) and the
antiproton spectra.

\section{Search for primordial antimatter}

The laws of fundamental physics being nearly the same for matter and antimatter, the picture of
a symmetric Universe made of a "patchwork" of regions of both types is very tempting. Theoretical
works \cite{cohen} have shown that, in such a model, gamma-rays induced by annihilation after 
recombination would exceed the observational limits. On the other hand, direct measurements
cannot exclude the presence of antimatter on scales larger than the typical size of galactic
clusters, around 20 Mpc. By trying to measure {\bf directly} anti-nuclei in cosmic rays, AMS
will be in position to strongly confirm the standard prediction or, possibly, to point out some
unknown phenomena. Nevertheless, whereas a positive detection (Z$>$2) would be clearly
conclusive, an upper limit on the amount of anti-nuclei in cosmic rays would be
difficult to be turned out into a lower limit on the domain size because of large
uncertainties on the structure and intensity of extragalactic magnetic fields.

Anyway, it should be noted that the experiment will also be very sensitive to
potential anti-globular clusters within our Galaxy \cite{belo}.

\section{Gamma-ray astrophysics}

Gamma-rays are not only interesting for astrophysical purposes (pulsars, SNRs, AGNs, GRBs, CIB
constraints, etc...) but also for the previously quoted search for neutralinos
annihilations \cite{Ber2}.

The AMS detector was not designed for gamma-ray astrophysics. In particular, it cannot be
pointed toward the sources. Nevertheless, together with AGILE and waiting
for GLAST, it could feature some interesting gamma-ray capabilities \cite{batiston}. Between 0.3
and 50 GeV, the angular resolution would be around $0.8^o(E/1GeV)^{-0.96}$ with a peak effective 
area of 1500 cm$^2$ and a flux sensitivity of the order of 0.5$10^{-8}$
cm$^{-2}$~s$^{-1}$~GeV$^{-1}$ (at 1 GeV). The calorimeter could also be very usefully used for 
gamma-ray astronomy if it can be self-triggered.

\section{AMS-I preliminary flight results}

Together with excellent technical results, accurate measurements of particles flux close to 
Earth have been performed by the AMS experiment, bringing a body of excellent new data on the 
particle populations in the low altitude terrestrial environment
\cite{ams1}~\cite{ams2}~\cite{ams3}~\cite{ams4}~\cite{ams5}~\cite{ams6}. Most of those measurements were quite surprising (high 
sub-geomagnetic cutoff population, $e^+/e^-$ ratio, $^3$He nuclei, etc...) and are now being
interpreted \cite{der1} \cite{der2}. They show that the detector worked very well and open a new
era of cosmic-ray physics.


\section*{References}
\begin{thebibliography}{99}

\bibitem{maurin} D. Maurin, F. Donato, R. Taillet \& P. Salati, accepted by ApJ (2001),
astro-ph/0101231
\bibitem{BO00} A. Bouchet, M. Bu\'enerd, L. Derome, Nucl. Phys. A, in press, (2001)
\bibitem{turner} M.S. Turner, Third Stromlo Symposium: The Galactic Halo, eds. B.K. Gibson, T.S. Axelrod, and M.E.
Putnam , Astron. Soc. Pac. Conf. Series, Vol. 666, (1999)
\bibitem{Bu} M. Bu\'enerd, XXIV Symposium on Nuclear Physics, Taxco, Mexico January 3-6, (2001)
\bibitem{simon} M. Simon, A. Molnar \& S. Roesler, ApJ, 499, 250, (1998)
\bibitem{Ber} L. Berstr\"{o}m, J. Edsj\"{o}, P. Ullio, ApJ, 526, 215, (1999)
\bibitem{donato} F. Donato, D. Maurin, P. Salati, A. Barrau , G. Boudoul \& R. Taillet submitted to ApJ,
astro-ph/0103150
\bibitem{donato2} F. Donato, N. Fornengo, P. Salati, Phys.Rev. D62, 043003, (2000)
\bibitem{jane} J.H. MacGibbon, B.J. Carr, ApJ, 371, 447, (1991)
\bibitem{orito} K. Maki, T. Mitsui, S. Orito, Phys.Rev.Lett. 76, 3474-3477, (1996)
\bibitem{cohen} A.G. Cohen, A. De Rujula, S.L. Glashow, ApJ, 495, 539 (1998)
\bibitem{belo} K. M. Belotsky, Yu. A. Golubkov, M. Yu. Khlopov , R. V. Konoplich \& A.S. Sakharov, INP MSU
98-31/532, (1988)
\bibitem{Ber2} L. Bergstrom, P. Ullio, J. Buckley, Astropart.Phys. 9, 137, (1998)
\bibitem{batiston} R. Battiston, Workshop on "GeV-TeV Gamma-Ray Astrophysics", Snowbird, Utah,
(1999), astro-ph/9911241
\bibitem{ams1} The AMS coll, Phys. Lett. B461 387, (1999)
\bibitem{ams2} The AMS coll, Phys. Lett. B472 215, (2000)
\bibitem{ams3} The AMS coll, Phys. Lett. B484 10, (2000)
\bibitem{ams4} The AMS coll, Phys. Lett. B495 440, (2000) 
\bibitem{ams5} The AMS coll, Phys. Lett. B490 27, (2000)
\bibitem{ams6} The AMS coll, Phys. Lett. B494 193, (2000)
\bibitem{der1} L. Derome, M. Bu\'enerd, A. Barrau, A. Bouchet, A.
Menchaca-Rocha \& T. Thuillier, Phys.Lett. B489, 1-8, (2000)
\bibitem{der2} L. Derome, M. Bu\'enerd, Y. Liu, submitted to Phys.Lett. B, (2001), astro-ph/0103474

\end{thebibliography}
\end{document}